# Document Archiving, Replication and Migration Container for Mobile Web Users

Peter Stanski, Stephen Giles and Arkady Zaslavsky

Department of Computer Technology, Monash University
900 Dandenong Road, Caulfield East, Vic 3145, Australia
{pstanski, sgiles, azaslavs}@broncho.ct.monash.edu.au

### Abstract

*With the increasing use of mobile workstations for a wide variety of tasks and associated information needs, and with many variations of available networks, access to data becomes a prime consideration. This paper discusses issues of workstation mobility and proposes a solution wherein the data structures are accessed in an encapsulated form - through the Portable File System (PFS) wrapper. The paper discusses an implementation of the Portable File System, highlighting the architecture and commenting upon performance of an experimental system. Although investigations have been focused upon mobile access of WWW documents, this technique could be applied to any mobile data access situation.*

## 1. Introduction

The mobile workstation such as Personal Digital Assistant (PDA) or notebook, have become indispensable tools for the traveling business person. Personal communication devices such as the Apple Newton, USRobotics/3COM PILOT, Nokia Communicator 9000, along with a plethora of PDAs running Windows CE, offer network connectivity with a limited bandwidth. These devices are often used to connect to the Internet for a variety of applications ranging from email to World Wide Web access.

Many organizations are appearing on the web daily. They offer general user access for advertising and marketing, while allowing internal staff to access more confidential information. The traveling business person often makes use of the web to provide product information via a wireless network. Wireless mobile computing is prone to dropouts and may offer limited coverage. This may occasionally leaves the user without network access, and force the browser to offer those documents which are in the cache.

Some users plan ahead, and web-copy documents from their organization's server to their mobile machine. A common solution involves replicating the necessary files along with their sub-directories. This procedure often leaves the user with a significant number of files scattered on local storage. To achieve this, the user has several ways to replicate a server's directory tree:

- Interactively follow every link while saving them to local storage (filenames need to be known in advance);
- Use an automated web-copy tool to duplicate the required web tree;
- FTP required documents to local storage (access pending).

HTML's open architecture and document building from distributed files is flexible but does not explicitly state file relationships until they are closely examined or viewed. A missing root document may introduce a severe user inconvenience if not duplicated.

In this paper we propose a new approach to web document management. We introduce the concept of a document wrapper which packages entire web servers sites or parts of their directory tree. Packaged documents are placed in a single file while permitting live access to the archived files. The following sections presents possible solutions to the replication problem and an alternative to disconnected operations. In later sections we discuss our project implementation, future extensions and related work.

## 2. Distributed File Systems Approach

The disconnected operations problem has already been addressed in some distributed file systems. Disconnection problems associated with Sun's NFS, the Andrew File System [1] and Ficus [6] lead to the development of the Coda file system at Carnegie Mellon University [3].

Ficus is a distributed UNIX file system developed at the University of California, Los Angeles. It employs a peer-to-peer file sharing model and offers multiple file replication. In cases of a node disconnection, the closest copy of the requested file is used. File modifications to a local copy are advertised to all other nodes once communication is re-established. Ficus differs from Coda in that it is a peer-to-peer system as opposed to a client-server environment.

Coda allows for disconnected operation by maintaining a small (approximately 50Mb) storage cache on each workstation. During disconnected periods, Coda would enter emulation mode and log local file access. Upon reconnection, Coda would enter reintegration mode, and resolve cached file modifications with the primary copy by playing though the log. Coda is an awkward scheme to implement as it requires significant modifications to an existing file system. Additionally, implementation of a Coda like file system caching into every client device may not always be possible or worthwhile.

A similar file system based on Coda, is the Hybrid File system for mobile workstations [7]. Developed at the University of Notre Dame. It combines ideas from Ficus and Coda, by allowing peer-to-peer file sharing and Coda like operation while disconnected from a file server. It is a superset to Coda, as it employs Coda's code base and extends it to provide peer sharing. Essentially, our needs for disconnected operations are addressed by Coda alone.

Using Coda may appear as an excessive approach. It requires a shared file system which differs from the operations of the Web. We propose this approach mainly because browsers are capable of opening locally stored files which may exist on remote mounts. This technique extends beyond the browser's operation and allows other applications to use these facilities while disconnected. Files located on the web server would appear as local therefore permitting applications to load selected documents.

### 2.1 Using Local Replicas

Currently companies and some universities deliver Web documents via a CD which is a much simpler approach then distributed file systems. The first generation of CDs provided documentation with a web like interface while the current releases are hybrids. These allow local browsing while maintaining absolute URL links to frequently changing information.

These solutions are often ideal but reflect associated costs of manufacturing. Another alternative is to provide local document copies in volatile storage. These are cheaper than CD-ROM mastering for modest sized document volumes.

For small sized replicas, their impact on local storage may not be significant. While large document replicas may substantially impact a file system. Since web documents are composed of several small files, these could potentially consume more disk space then required because of i-node or cluster sizes.

### 3. Link Related Issues

Most links within documents are relative, meaning that images embedded in the document are located relative to the HTML's location. Duplicating these documents do not present a problem, provided the directory structures are maintained.

While absolute link addressing is used to point a browser at a remote site as it specifies a whole URL. Many web authors use absolute links to their own site and unknowingly create documents which are difficult to copy. Such absolute links present a problem for simple document replication. They also force the browser to obtain documents via HTTP which breaks operations while opening local files. It should be noted that most web-copy tools, are intelligent and replace absolute links with relative links reflecting the documents new location. Such tools significantly aid in web authoring and replication.

Another alternative is to adopt an authoring standard for a site where all documents use relative links.

### 4. Document Archiving and Storage

Many users are familiar with archiving utilities such as pkzip and tar. These collect files and put them into a single package. Currently there are no tools which allow web document archival and easy retrieval. It is possible to use the above mentioned tools but until now, servers did not have hooks into these programs to provide seamless integration. Through the usage of a packaging tool it would be possible to combine all the necessary web documents into a single file. This would make document distribution and replication much easier for a user and permit administrators to archive aging documents.

We propose the packaging approach as a solution to simplify web document replication, minimize the impact on a local file system and assist in site archival while maintaining its web document presence.

### 4.1 Web Document Wrapper

The web document wrapper named the Portable File System (PFS), has been developed as a part of the ongoing PESOS project at Monash University [8]. PFS originated as a container for collecting mobile agent component files into a single migration archive. Combining component into one file reduced migration complexity and allowed for greater manageability.

In the case of HTML documents, the PFS contents are those of web documents duplicated from an existing web server. Figure 1 below, shows the PFS contents for a simple directory tree. Through the use of PFS wrappers, whole or partial directory trees may be archived, become organized by collapsing subdirectories while maintaining their contents intact. In the case of Figure 1, a single PFS file could replace all existing files in a given Accounts directory along with Images, Dept1 and Dept2 subdirectories. This example reflects the embeddable and remote referencing facilities offered by a PFS wrapper.

Traditional file archiving tools have relied on embedding all files within the archive file itself. Our extended approach to archiving allows for documents to be remotely referenced using standard Internet protocols. Figure 1 depicts two image files as being stored at the ABC.COM site, which may be obtained using HTTP or FTP protocols.

For majority of users remote file referencing will not be necessary as they shall require remote files to always be available. In cases where a user wishes to sacrifice content for availability and size, it would be possible for the archive to contain embedded textual information while all optional files (images) would be remotely tagged within the PFS. In this scenario, users experiencing network dropouts would only access locally available HTML documents without obtaining the remote graphical content.

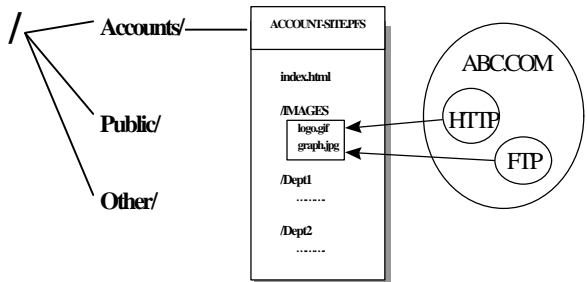

**Figure 1: Directory structure within a PFS wrapper**

### 4.1 Internal PFS Structure

To maintain compatibility with future PFS revisions, the internal structure has been kept open. This permits future wrapper improvements while maintaining backward compatibility. To achieve open structure the file definitions and storage are a hybrid ASCII-Binary format. All embedded files may be stored as uuencoded/raw-binary contents, while file descriptions are described in plain text. Textual descriptions allow encoding/decoding code to interpret tags which have semantics relevant to a software revision.

A sample PFS description for a remotely tagged file is given below along with the wrapper header:

```
PFS!
version=1.0
date=25-06-97
.. .. ..
[ENTITY]
originalname=C:\Program Files\WINZIP\Vendor.txt
longname=Vendor.txt
shortname=Vendor.txt
dirname=                  ;PFS directory (root)
created=14-08-95 6:00:00 AM
length=2952   ;physical file length
origin=Windows95          ;created Win-PFSutil ver1.0
description=This is the Vendors message file
remotereadhost=http://astral.ct.monash.edu.au/~files/vendor.txt
mode=RO                   ;read-only access
storage=remote            ;stored on remotehost
.. .. ..
[ENTITIY]
```

### 5. Accessing PFS Contents

In order to make use of PFS wrappers and directory collapsing, small changes have to be introduced to the way web servers resolve a URL specifying a PFS path.

A HTTP client issuing a *get* request, specifies the file to be transferred. In the case of Figure 2, the URL incorporates a HTML file to be found within a PFS. Under normal circumstances, a server is required to establish the physical path from the URL. In this case it would result in a "404 Error: not found" when processed by a standard server. This is assuming that a directory does not exist with an identical name as the PFS.

```
http://broncho.ct.monash.edu.au/~pstanski/account-site.pfs/index.html
                                                    PFS filename   path/filename
```

**Figure 2: Format of a GET URL with a PFS reference**

Our prototype server parses the URL and establishes if the requested file is a part of a PFS package. In the case of the above URL, *"index.html"* would be searched inside the *"account-site.pfs"* file. If found, the file would be sent to the client, or fail with an error such as file not found. The parsing modifications make our implementation transparent to the client user.

### 5.1 Obtaining PFS Packages

In order to obtain PFS packages which contain a web server's directory tree, a client may request the wrapper file as specified in Figure 3. The appropriate MIME type is set to allow for a binary file transfer to the client's browser.

```
http://broncho.ct.monash.edu.au/~pstanski/account-site.pfs
                                                    PFS filename
```

**Figure 3: URL to obtain the PFS wrapper file**

Prior to our PFS server extensions, users had two ways of obtaining web documents via HTTP. They could download each page and its embedded images by hand, or use an automated tool which would do it for them. PFS now allows for obtaining archived documents and to view them seamlessly.

### 6. Internal PFS Access and Implementation

Access within a PFS package may be implemented in a variety of ways. The following briefly describes some possible alternatives:

*As an Archive:*

The package could be viewed and have the documents extracted by a PFS management utility (pfsutil). This is identical to accessing a tar or zip archive. A received PFS could be extracted with directory entries, and browsed locally. Files tagged as remote and maintained on a central server, could seamlessly become available for users as soon as they became available on the server.

*As a File System:*

Through the development of an installable file system, a PFS file could be mounted as a volume in a UNIX environment. In the case of DOS/Windows, a RAM drive could provide access. This is perhaps the most flexible approach, allowing any application to browse a mounted document package. The remote tagging of files would create a read-only distributed file system.

*Within a Proxy:*

A HTTP proxy could be extended to provide advanced caching facilities for an entire web site. PFS creators would have the possibility of maintaining static information as embedded files while frequently changing file contents could be tagged as remote files. Web clients using a proxy would only need to obtain remote files as needed, with embedded files being returned by the local proxy.

*Within a Server:*

Provide browsing and document retrieval as described in the previous sections.

In our implementation we opted for the server option. We used a freely available web server called Boa (v0.92). Originally developed by Paul Philips as it appeared suitable and easily extendible. Boa+PFS as it is now called, runs under the Linux 2.0 operating system. Figure 4.0, provides an overview of our implemented architecture.

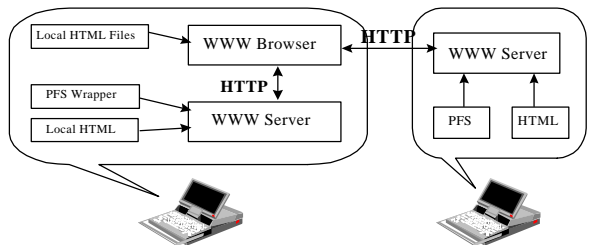

**Fig.4: Local implementation of PFS handling is done by Boa WWW server**

PFS files held locally are accessed through a browser by specifying their local URL. These require appropriate permissions to be set in order to be accessible by Boa. The local Boa server, browses the specified file and returns documents held within them. Additionally, a remote Boa server may be used to provide the same functionality.

## 6.1 Server Performance

New additions to servers often impact on their performance. Therefore, modifications to file handling and retrieval algorithms are likely to increase the current server's processing overheads. In our case we tested server performance on an CSMA/CD (ethernet) LAN, with the server and client workstation. The server was a Pentium 120Mhz notebook and the client a 486DX/100Mhz desktop system. The following results were obtained for the original version of the server (v0.92) and the PFS compliant server (v0.93):

|  | *Access to a normal page* | *Access to page within PFS* |
|---|---|---|
| *Original Server* | *2.5 sec* | *---* |
| *PFS Compliant Server* | *2.5 sec* | *2.5 sec* |

**Table 1: Access times for web page- 3 files using 26180 bytes in total**

We ran the test 10 times and obtained the following average values. In our experience, we found no reduction in the server's performance. Accessing documents within the PFS wrapper as opposed to non-wrapped documents yields the same results. There is in fact an overhead in code, but it does not impact on server performance.

In our implementation, we use a linear file searching algorithm which is not noticeable for small PFS containers. We suspect that the server may slow down and wrapper access may become noticeable for large packages containing hundreds of files. Figure 5, shows the test page upon which we based our timings. It shows a browser looking inside a sample page which is held within a remote "diet.pfs" package.

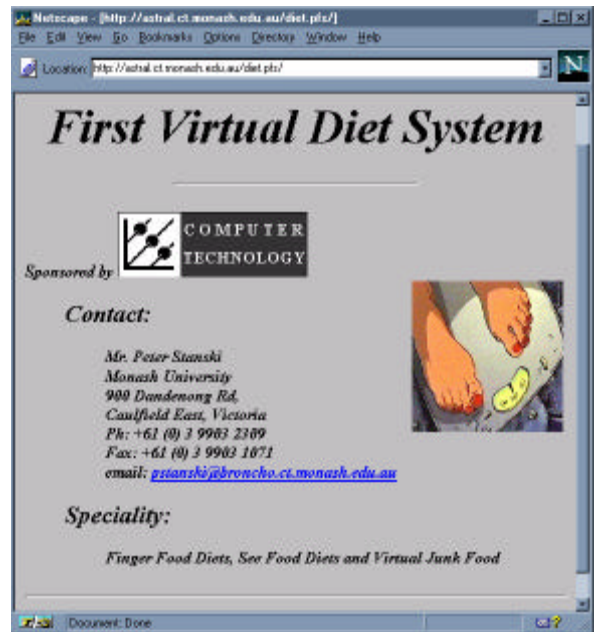

**Figure 5. Sample "DIET.PFS" wrapper used for performance testing**

## 6.2 Future Extensions

Future implementations and extensions, may provide a faster directory look up structure instead of the current linear search. We are extending the internal PFS file format to allow for future format extensions while maintaining backwards compatibility. This is likely to results in further BOA+PFS server modifications. Additional work in developing an installable file system for Linux/Windows95 is also being examined.

## 7. Related Work

Distributed wireless multimedia access has gained significant attention over the last several years. Relevant projects are listed below to provide the reader with some further insight into the field.

The Defense Advanced Research Project Agency (DARPA) has initiated the Global Mobile Information Systems (GloMo) program in 1994 [5]. It is a testbed initiative linking several research projects which depend on a wireless communications infrastructure. One such project at the University of California, Berkeley, is the lightweight, portable multimedia notebook called InfoPad [4]. It aims to provide multimedia access in a distributed wireless environment. InfoPad performs as a mobile terminal which maintains applications on dedicated servers. All user applications execute remotely, while their results are sent to the InfoPad terminal.

Web oriented projects such as Mobisaic and Mowser are indirectly related to our wrapper project. The Mobisaic [9] project at the Univeristy of Washington, employs context sensitive URLs in a wireless local area network to provide location dependent web documents. We also employ modified URLs while leaving the server to resolve them. A mobile Mobisaic user communicates with physically visible devices via infra-red to obtain environmental variables. These are then incorporated within dynamic URLs which provide the user's location by retrieving the corresponding web pages. Mobisaic proposes the concept of browser registration and permits server callbacks to client browsers. Browser callbacks assist in document updates by notifying registered browsers to obtain newly changed documents.

Another multimedia project called Mowser, has been developed at Purdue University [2]. The Mowser architecture employs designated proxy servers which service mobile users. These examine incoming documents while caching them. A Mowser proxy modifies retrieved documents and forwards resized images to mobile browsers. This reduces the necessary bandwidth requirements while delivering the document content.

The University of Washington, developed the W* wireless web browser [10]. Their system employed a low powered CPU in a wireless Personal Digital Assistant (PDA). This appears to be similar to Mowser as both projects employ proxy servers to assist document modification and preprocessing. The mobile PDA running W* communicates with a specialized proxy server via a compressed protocol. The proxy caches previously accessed documents, preprocesses browser displays, prefetches the first page of potential next links, and passes compressed documents via the wireless connection.

To the best of our knowledge, document management such as the PFS wrapper approach has not been attempted and appears to be a new contribution to this field of work.

## 8. Conclusion

The PFS wrapper approach to archiving sites while maintaining their web presence is a new server feature. It permits site administrators to collapse directories and offers packaged files to users wishing to access portions of a web server tree. Additionally, the support for remote file referencing without physically embedding files have been used in distributed file caching systems but appears to be a novel approach for archives. The PFS code addition within our server, provides minimal overhead for document access and does not degrade server performance.

Mobile users which are capable of running a small local web server, may access PFS files and retrieve necessary documents while disconnected from a network. The use of a single PFS wrapper minimizes the need to potentially maintain hundreds of files on local storage. The portable file system also simplifies multi-document distribution. It seamlessly integrates into a web server and appears transparent to its users.

**Acknowledgments**

The authors would like to thank John Ng for obtaining the original Boa archive and the inspiration which led to the extension of Boa.